\documentclass[%
 reprint,
 superscriptaddress,
 showpacs,preprintnumbers,
 bibnotes,
 amsmath,amssymb,
 aps,
 pra,
 floatfix,
]{revtex4-1}

\usepackage{graphicx}
\usepackage{subfigure}
\usepackage{color}
\usepackage{dcolumn}
\usepackage{bm}
\usepackage{amsmath}

\begin{document}


\title{Continuous-time quantum walk spatial search on the Bollob\'{a}s scale-free network}

\author{Tomo Osada}
\email{tosada92@gmail.com}
\affiliation{Tokyo University of Science, 1-3 Kagurazaka, Shinjuku-ku, Tokyo 162-8601, Japan.}

\author{Bruno Coutinho}
\affiliation{Instituto de Telecomunica\c{c}\~oes, Physics of Information and Quantum Technologies Group, Portugal.}

\author{Yasser Omar}
\affiliation{Instituto de Telecomunica\c{c}\~oes, Physics of Information and Quantum Technologies Group, Portugal.}
\affiliation{Instituto Superior T\'{e}cnico, Universidade de Lisboa, Portugal.}

\author{Kaoru Sanaka}
\affiliation{Tokyo University of Science, 1-3 Kagurazaka, Shinjuku-ku, Tokyo 162-8601, Japan.}

\author{William J. Munro}
\affiliation{National Institute of Informatics, 2-1-2 Hitotsubashi, Chiyoda-ku, Tokyo 101-8430, Japan.}
\affiliation{NTT Basic Research Laboratories \& NTT Research Center for Theoretical Quantum Physics, NTT Corporation, 3-1 Morinosato-Wakamiya, Atsugi, Kanagawa 243-0198, Japan.}
\author{Kae Nemoto}
\affiliation{National Institute of Informatics, 2-1-2 Hitotsubashi, Chiyoda-ku, Tokyo 101-8430, Japan.}
\textbf{•}

\date{November 9, 2019}

\begin{abstract}
The scale-free property emerges in various real-world networks and is an essential property which characterizes the dynamics or features of such networks. In this work we investigate the effect of this scale-free property on a quantum information processing task of finding a marked node in the network, known as the quantum spatial search. We analyze the quantum spatial search algorithm using continuous-time quantum walk on the Bollob\'{a}s network, and evaluate the time $T$ to localize the quantum walker on the marked node starting from an unbiased initial state. Our main finding is that $T$ is determined by the global structure around the marked node, while some local information of the marked node such as degree does not identify $T$. We discuss this by examining the correlation between $T$ and some centrality measures of the network, and show that the closeness centrality of the marked node is highly correlated with $T$. We also characterize the distribution of $T$ by marking different nodes in the network, which displays a multi-mode lognormal distribution. Especially on the Bollob\'{a}s network, $T$ is magnitude of orders shorter depending whether the marked node is adjacent to the largest degree hub node or not. However, as $T$ depends on the property of the marked node, one requires some amount of prior knowledge about such property of the marked node in order to identify the optimal time to measure the quantum walker and achieve fast search. These results indicate that the existence of the hub node in the scale-free network is playing a crucial role on the quantum spatial search.

\end{abstract}

\maketitle


\section{\label{sec:1}Introduction} 
\begin{figure}[t]
 \centering
 \includegraphics[width=8cm]{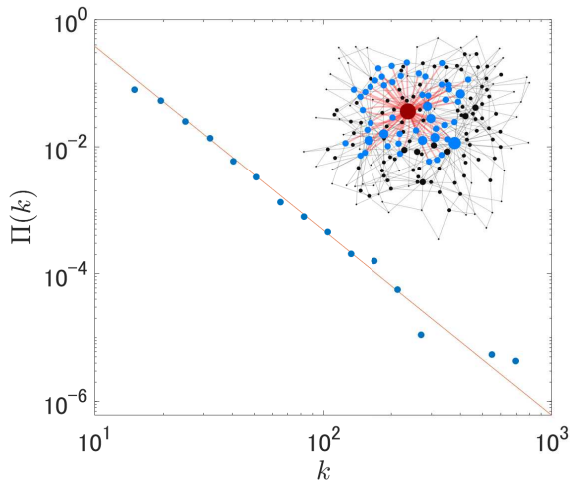}
 \caption{Degree distribution of the Bollob\'{a}s model with network parameters $N=10000, m=5, \beta=2.9$. The red solid line is the fitting curve of the blue data points acquired from the generated network. Inset: visualization of the network for $N=200, m=2, \beta=3$. The size of the nodes are determined by the closeness centrality of its node, the red node has the largest degree (the hub node) with its edges colored in red, and the blue nodes are the neighbours of the hub node.}
 \label{fig:degreedist}
 \end{figure}

 Social, technological or biological systems in the real-world often display complex interactions between each elements which cannot be simply explained as regular or random structures. Such real-world systems can be analyzed by mapping their interactions as a graph, often referred to as complex networks. Understanding the structural properties or simulating dynamics on these networks has revealed universal properties of real-world systems \cite{CNreviewStructure,CNreviewCritical,community}. Especially, scale-free networks are important class of networks as they commonly emerge in various systems, such as the World Wide Web, protein interaction in biological organisms, or transport systems like the airline network \cite{emergence,biological}. Scale-free networks are characterized by their degree distribution following a power law function of the form
 \begin{equation}
\Pi(k) \propto k^{-\beta},
\label{Eq:deg dist}
\end{equation}
where $k$ is the degree of a node and the exponent $\beta > 0$ is a real constant (see Figure \ref{fig:degreedist}). Simulating various dynamics on such networks led to comprehensive understanding of dynamics in real-world systems such as fast spreading of information \cite{CNreviewDynamics,EpidemicSpread,RumorSpread,epidemic}.

 On the other hand, recent development in engineering quantum systems has enabled us to design complex quantum systems that are beyond regular lattice structures \cite{MBL,waveguides,NVnetwork,junction}. Analyzing quantum dynamics or performing quantum information processing task on such complex quantum networks is becoming a great interest, and it is important to examine what difference or improvement we can see from these systems compared to regular lattice systems \cite{QuantumCommunity,interdisciplinary,QuantumComplex,QuantumInternet}.
 
 In this scope, here we analyze a quantum information processing task to find a marked node in a graph, which is known as the spatial search, on a scale-free network. Searching a marked item in a database is one of the most fundamental and important computational problem. The spatial search is the case where each item of the database is represented as each node of a network, and one aims to find a marked node using some quantum dynamics and measurement defined on the network \cite{Benioff,Ambainis,Childs}. One can achieve this by using the framework presented by Childs and Goldstone \cite{Childs}, which prepares a black box Hamiltonian that encodes the network adjacency matrix and the information of the marked node, and perform a continuous-time quantum walk using that Hamiltonian. Since the network structure is encoded in the Hamiltonian, we can observe how the scale-free property of the network will affect the quantum walk and the spatial search.
 
 Under the framework of continuous-time quantum walk, spatial search has been extensively studied on various graphs. Much work has been done considering regular graphs or lattice structures \cite{Childs,Childs2,graphene,Wong1,reduction}, as well as comprehensive analysis of spatial search using Erd\"{o}s R\'{e}nyi random graphs \cite{Chak1} or general Markov chains \cite{Chak2}. The main focus of these work were whether one can achieve the search with the time complexity of $O(\sqrt{N})$ on the given graph. Some work moved toward exploring non-regular structures. Agliari \textit{et al.} \cite{pra82.012305} explored spatial search on fractal structures and studied how the transition in the ground state of the Hamiltonian depends on the marked node. Berry and Wang \cite{pra82.042333} studied spatial search by discrete-time quantum walk on a Cayley tree, and examined the relation between the search and centrality measures of the graph. Philipp \textit{et al.} \cite{pra93.032305} examined on balanced trees, and derived that the search performance changes depending whether the marked node is towards the root or the leaves of the graph. Although each of these work provided important results to characterize some correlation between the graph structure and the spatial search, we still lack the knowledge how the spatial search will behave on complex networks. 

To further clarify the our aim, we point out two differences we expect between the complex networks we are interested in and the handful of graphs mentioned above. First, the nodes in a complex network are mostly non-equivalent to each other. A counter example is the nodes in lattices with periodic boundary condition, which are all equivalent due to the translational symmetry. Secondly, complex networks are not purely random and some order lives in the randomness. A typical example is the scale-free network, as they have hub nodes which have substantially large degree than the others, while most nodes having small degree. To satisfy these conditions, we choose the Bollob\'{a}s model \cite{bollobas1,bollobas2}, a mathematical model to generate the scale-free network obeying preferential attachment \cite{emergence}, as the network where to analyze the spatial search.

Searching nodes on scale-free networks using classical random walks have been investigated in the literature, in terms of analyzing the hitting time or the mean first passage time \cite{Noh2004,Bollt2005,Zhang2009}. On the Barab\'{a}si-Albert network, the mean first passage time is shown to be roughly proportional to the degree of the target node \cite{Noh2004}. Regarding the dependency on the network size $N$, the mean first passage time scales linearly to $N$ \cite{Bollt2005}, while sub-linear scaling were also found at special cases such as searching the hub node \cite{Zhang2009}. In our work, we will also discuss how the quantum nature leads to different results compared to the above classical cases.

 Through our numerical simulations, first we will show the speed of the spatial search using quantum walk indeed depends on which node in the network is marked, due to the non-equivalence of the nodes. Surprisingly, this speed can be different up to few orders of magnitude. This is the first critical difference compared to searching on regular or lattice graphs. We characterize how this dependency emerges in terms of the leading eigenvector (the eigenvector corresponding to the largest eigenvalue) of the network adjacency matrix. This reveals that the performance of the algorithm is dominated by the localized property of the leading eigenvector. To further characterize the relation between the network structure and the performance of the search, we examine the correlation between some centrality measures of the network and the time complexity of the quantum search. We find that the degree and the time complexity of the search is not strongly correlated, unlike the search using classical random walk. The speed is rather determined by the shortest paths distances between the marked node and the rest of the nodes. This observation cannot be seen from purely random graphs \cite{Chak1}, and this is another critical difference from the previous studies. We also point out one advantage of using a scale-free network for the spatial search, which is that one can perform the search starting from a localized initial state instead of a global superposition state conventionally used in spatial search. From this, one can naturally translate the spatial search to a efficient state transfer protocol between the hub node and another arbitrary node. All these results indicates that the hub node is playing an important role for the spatial search algorithm.

\section{\label{sec:2}Models}
 Let us begin by defining the spatial search algorithm we are going to examine. Defining $G(V,E)$ as a graph (such as shown in Figure \ref{fig:degreedist} inset) with a set of nodes $V = \{1,2,\ldots,i,\ldots,N\}$ and a set of edges $E$, we consider an $N$-dimensional Hilbert space spanned by the basis states $\{|1\rangle,|2\rangle,\ldots,|i\rangle,\ldots,|N\rangle\}$. Each state corresponds to the situation where a quantum walker is localized at node $i$. We can then define the state of the quantum walker at time $t$ as $|\psi(t)\rangle = \sum_{i=1}^{N} c_i |i\rangle$ with the $c_i$ constrained such that $\sum_{i=1}^{N} { |c_i| }^2=1$. In order to search for a single marked node (labelled $|w\rangle$ in this case), we let the state $|\psi(t=0)\rangle$ evolve under the action of the Hamiltonian
 \begin{align}
H &= -\gamma A - \epsilon_w |w\rangle\langle w|\\
  &= -\gamma \sum_{i,j}^{N} A_{ij} \left( |i\rangle\langle j| + |j\rangle\langle i| \right) - \epsilon_w|w\rangle\langle w|,
\label{Eq:Hamil}
\end{align}
where $A$ is the adjacency matrix of graph $G(V,E)$ whose entries are defined as $A_{ij} = A_{ji} = 1$ if nodes $i$ and $j$ are connected by an edge, and otherwise $A_{ij} = 0$. The real constant $\gamma \geq 0$ is the transition energy between the nodes and $\epsilon_w$ is the on site energy on node $w$. The projection $|w\rangle\langle w|$ causes the amplitude to accumulate on the marked node $w$. We consider the unitary time evolution of the system given by
\begin{equation}
 i\hbar \frac{d}{dt} |\psi(t)\rangle = H|\psi(t)\rangle,
\label{Schrodinger}
\end{equation}
 and compute the probability to measure the quantum walker at the marked node $P(t) = |\langle w|\exp{({-i} H t /\hbar)}|\psi(0) \rangle|^2$. The time complexity of the algorithm or the ``search time'' $T$, in units of $\epsilon_w$, is evaluated by finding the shortest time $t = \tau$ that maximizes $P(t)$. As one can find the quantum walker on node $w$ with probability $P(\tau)$ at the optimal measurement time $\tau$, the algorithm can identify the marked node with success probability $P(\tau)$. We finally compute $T = \tau/P(\tau)$ which takes into account the repetition of the algorithm for $1/P(\tau)$ times.
 
 Although we are aware that the Hamiltonian in Eq. (\ref{Eq:Hamil}) has limitations when the underlying graph is non-regular \cite{Glos1}, and a modified search Hamiltonian has been proposed \cite{Chak1}, we use the Childs and Goldstone's formalism since our interest is focused on investigating how the quantum dynamics is affected when the Hamiltonian itself has a scale-free property. Modifying the Hamiltonian based on the method by Chakraborty \textit{et al.} \cite{Chak1} will be advantageous to analyze the time complexity on arbitrary graphs systematically, but also compensates the inhomogeneity of the graph. This will compensate the scale-free property of the underlying graph, which contradicts with the purpose of this paper. Additionally, the search Hamiltonian by Childs and Goldstone could be experimentally realized on quantum simulators \cite{singlemagnet}, without requiring gate decomposition of the algorithm on quantum computers. For these reasons, in our work we keep our focus on the Hamiltonian formalized by Childs and Goldstone.

 Next we describe how the preferential attachment (PA) model is generated and point out some properties of this network. We use the formalization by Bollobas \cite{bollobas1,bollobas2}. The process of generating the network with $N$ nodes is as follows: At the initial time step $u=1$, the network $G_{\lbrace u = 1 \rbrace}$ starts with a single node $v'_1$ with one edge connecting to itself. At every subsequent time step $u \geq 2$, one node $v'_{u}$ having one outgoing edge is added into $G_{\lbrace u-1 \rbrace}$ and connects to one of the nodes in $G_{\lbrace u \rbrace}$ with its outgoing edge. Defining the degree of node $v'_i$ at time $u$ as $d_{u}(v'_i)$, the node to connect to is chosen by the following probability distribution \cite{bollobas1}
 
 \begin{equation}
 Pr (i=s) =
  \begin{cases}
  d_{u-1}(v'_s) / (2u-1) & 1 \leq s \leq u - 1 \\
  1 / (2u-1) & s=u.
  \end{cases}
\label{Eq:preferential}
 \end{equation}
This means that the probability for a node to be chosen is proportional to its degree, which resembles the ``preferential attachment''. After repeating the above process until a cetain time step $u = m \quad (m \in \mathbb{N})$, the set of nodes $v'_{1}, v'_{2}, \cdots, v'_{m}$ forms a single node $v_1$. The edges that were connecting the nodes within the set is converted to $m$ self loops on $v_1$. The process of adding new nodes $v'_{u}$ is continued until time step $u=2m$, and again the the set of nodes $v'_{m+1}, \cdots, v'_{2m}$ forms another node $v_2$. If $m' \leq m$ nodes in the set of nodes $v'_{m+1}, \cdots, v'_{2m}$ are connected to $v_1$, they are converted to $m'$ edges between $v_1$ and $v_2$. Following the rule described above, the process is repeated until time step $u_{\text{end}} = mN$, which results as a network $G_{\lbrace mN \rbrace}$ with $N$ nodes and $mN$ edges. The obtained network has a power law degree distribution with exponent $\beta=3$ \cite{bollobas1}. In order to change the value of $\beta$, we use the method introduced by Dorogovtsev \textit{et. al} \cite{BAvar}.

From the construction above, we have three control parameters when generating the network; the total number of nodes $N$, the parameter which controls the connectivity of the network $m$, and the degree distribution exponent $\beta$. The average degree of the network is $2m$, while the minimum degree is $m$ and the largest degree is $\sim N^{1/(\beta-1)}$. Note that we allow self loops and parallel edges between nodes in our network in order to keep consistency, that is to fix the total number of edges to $mN$ for every trial of generating $G_{\lbrace mN \rbrace}$. When converting $G_{\lbrace mN \rbrace}$ to the adjacency matrix $A$, the number of self loops or parallel edges contributes to the weight of the diagonal or the off-diagonal entries of $A$, respectively. The degree distribution and visualization of an instance of $G_{\lbrace mN \rbrace}$ is shown in Figure \ref{fig:degreedist}. 

Although there are many other scale-free network models proposed in the literature \cite{fitness,hierarchical,generation,fractal,hyperbolic}, in this paper we focus only on this Bollob\'{a}s model to make our problem more tractable. The Bollob\'{a}s model has no high clustering coefficient, community structure, or a self-similar structure. We leave the examination of the effect of such properties on the spatial search for future work, and take advantage of the simplicity of Bollob\'{a}s model to concentrate on how the power-law degree distribution affects the spatial search.

\section{\label{sec:3}Search time on the Bollob\'{a}s model}

  \begin{figure}[htp]
 \centering
 \includegraphics[width=8cm]{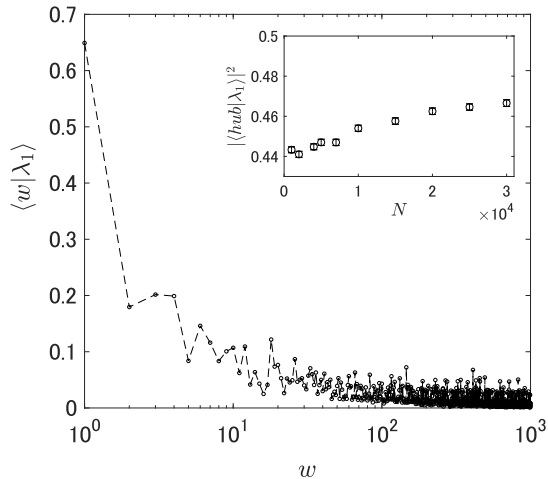}
 \caption{Plot of the overlap $\langle w | \lambda_1 \rangle$ verses $w$, showing all components of the leading eigenvector (the eigenvector corresponding to the largest eigenvalue of the network adjacency matrix) from one instance of our Bollob\'{a}s model with network parameters $N=1000, m=10, \beta=3$. The eigenvector, which all components are real and positive, displays a localized property around the large degree nodes. Index $w=1$ is the largest degree hub node. Inset: The squared component on the largest degree node $|\langle \text{hub} | \lambda_1 \rangle|^2$ for different sizes of the Bollob\'{a}s model with $m=10$ and $\beta=3$. Since there is a substantial overlap between $|\text{hub}\rangle$ and $|\lambda_1\rangle$, and the value does not decrease when $N$ is increased, we can select the state $|\text{hub}\rangle$ as the initial state without degrading the scaling of the search time.}
  \label{fig:eigenvector}
 \end{figure}
 
\begin{figure}[t]
 \centering
 \subfigure[]{\label{fig:opt-1}
  \includegraphics[width=7.2cm]{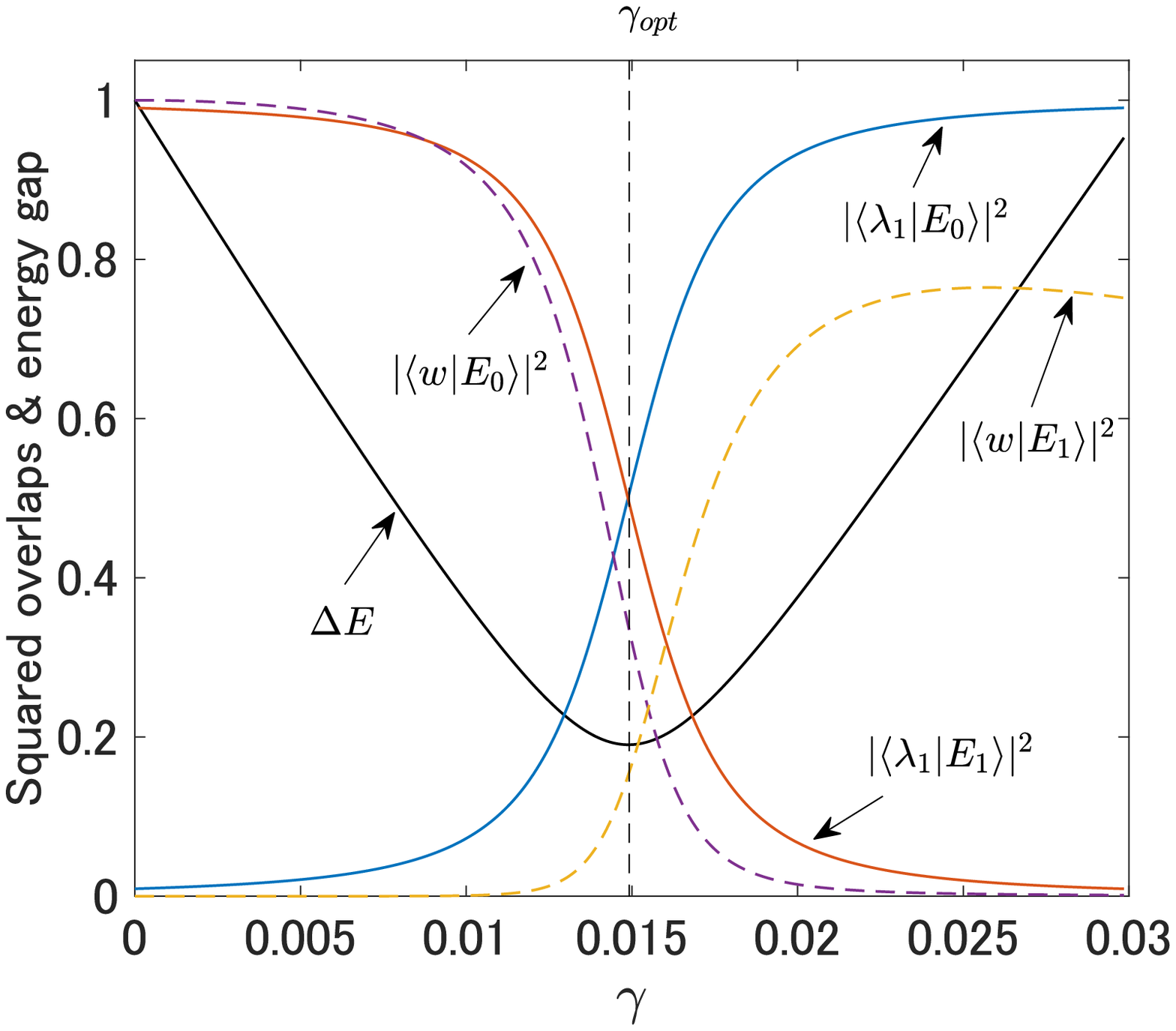}}
 \subfigure[]{\label{fig:opt-2}
  \includegraphics[width=7.2cm]{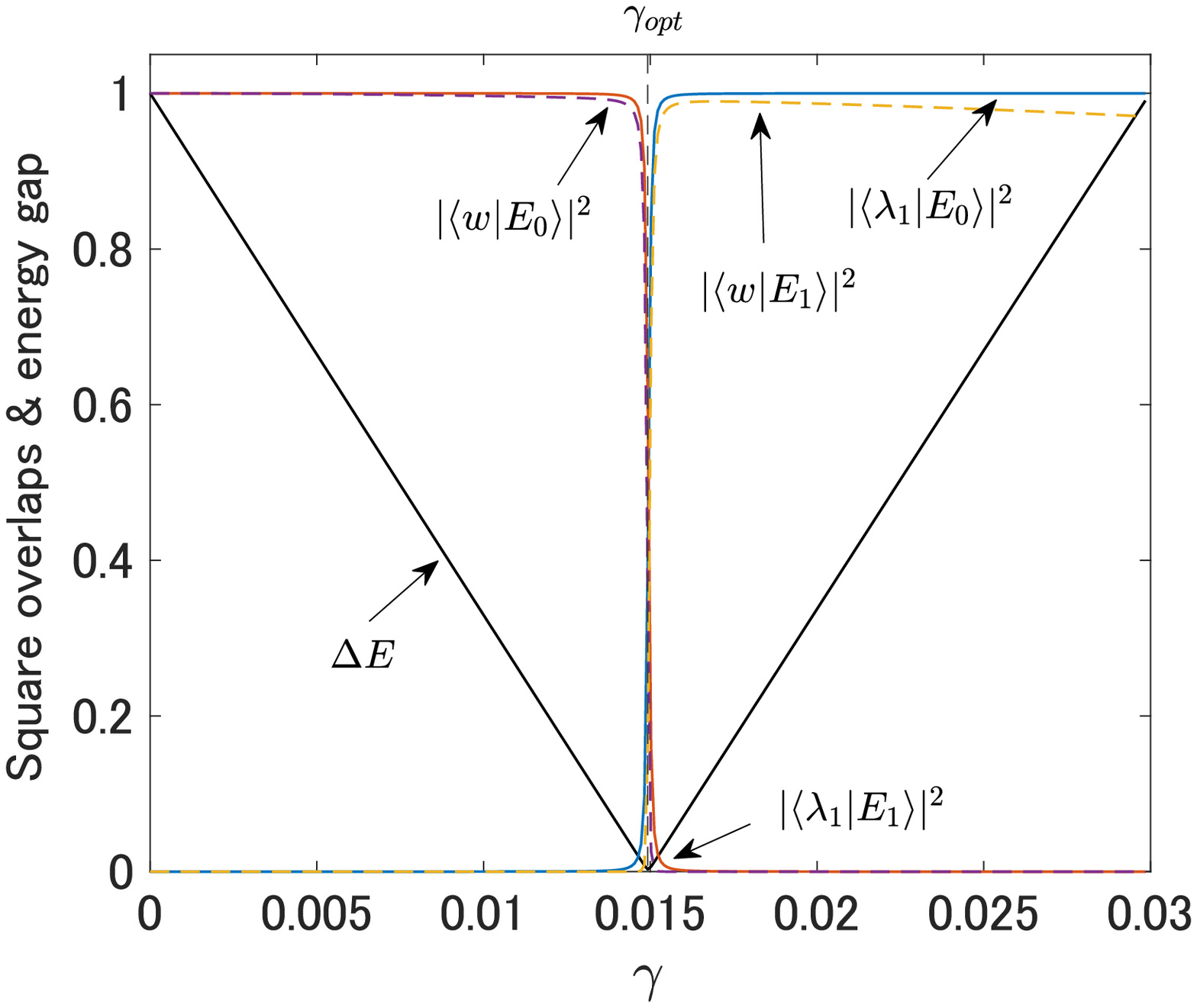}}
  
  \caption{Plots of the squared overlaps of the states of interest, and the energy gap $\Delta E \equiv E_1 - E_0$ of the Hamiltonian $H$. The Bollob\'{a}s model with $N=2000, m=10, \beta=3$ is used in this plot. By changing the value of $\gamma$, we can see the quantities $|\langle \lambda_1|E_{0,1} \rangle|^2$ changes from $0$ to $1$ and vice versa, which confirms that the eigenstates $|E_{0,1}\rangle$ switches around $\gamma = \gamma_{opt}$. This also confirms that $|\lambda_1\rangle = (|E_{0}\rangle - |E_{1}\rangle)/\sqrt{2}$ is approximately achieved at $\gamma = \gamma_{opt}$. $\Delta E$ is minimized at this point. The quantities $|\langle w|E_{0,1} \rangle|^2$ are shown to indicate how close the resulting state of the time evolution $|\tilde{w}\rangle$ is close to $|w\rangle$. Figure (a) represents the case when a node with a large degree is marked, and (b) represents the case when a node with smallest degree in the network is marked. (a) has the larger $\Delta E$ at $\gamma = \gamma_{opt}$ compared to (b), which indicates that the evolution time $\tau$ is smaller. One can also see the switch of eigenstates around $\gamma = \gamma_{opt}$ is much sharper in (b).}
  \label{fig:opt}
 \end{figure}

 As the first step to analyze the search time on the Bollob\'{a}s model, we consider an abstract dynamics of the spatial search algorithm to show that the search time will depend on the selection of the target node in the network. 
 We define two states $|\lambda_1 \rangle$ and $|\tilde{w} \rangle$ as
\begin{align}
 \label{eq:stateE0}
|\lambda_1 \rangle &= (|E_0\rangle - |E_1\rangle)/\sqrt{2}, \\
 \label{eq:statewt}
|\tilde{w} \rangle &= (|E_0\rangle + |E_1\rangle)/\sqrt{2},
\end{align}
where $|E_{0,1} \rangle$ are the lowest and second lowest energy eigenstates of our Hamiltonian $H$, with the parameter $\gamma$ chosen at a specific value $\gamma = \gamma_{opt}$. We assume that the two lowest energies are non-degenerate. We also assume that $|\lambda_1 \rangle$ is the leading eigenvector of $A$, such that $A|\lambda_1 \rangle = \lambda_1|\lambda_1 \rangle$. This assumption of Eq. (\ref{eq:stateE0},\ref{eq:statewt}) is based on degenerate perturbation theory \cite{Childs}, and can be also confirmed from Figure \ref{fig:opt}. Next we are going consider the unitary evolution where the initial state is $|\lambda_1 \rangle$, which will rotate to $|\tilde{w} \rangle$ in time $\pi/\Delta E$. Here $\Delta E \equiv E_1 - E_0$ is the gap between the energies corresponding to the eigenstates $|E_{0,1} \rangle$. It is straightforward to show that
\begin{align}
 \label{eq:energygap}
\Delta E = 2|\langle \lambda_1|w \rangle \langle w |\tilde{w} \rangle|.
\end{align}
The first factor in Eq. (\ref{eq:energygap}) tells us that the energy gap (and equivalently the evolution time $\tau = \pi / \Delta E$) depends on the component $c_w$ of the leading eigenvector $|\lambda_1\rangle = \sum_{i=1}^{N} c_i |i\rangle$. One has $\langle \lambda_1|w \rangle = 1/\sqrt{N}$ for any index $w$ if the adjacency matrix $A$ is the one for regular graphs or if the Laplacian matrix is used. However, for non-regular graphs the components of the leading eigenvector is not uniform. As in the preferential attachment network, it was shown by Goh \textit{et al.} \cite{eigenvector} that the components of $|\lambda_1 \rangle$ is localized on the largest degree node, and $c_i$ varies from $1/\sqrt{2}$ to $1/(2\sqrt{N})$. Figure \ref{fig:eigenvector} confirms this property of $|\lambda_1\rangle$ for the network we have generated. The value of the second factor in Eq. (\ref{eq:energygap}) is non-trivial, since we need to know $\langle w|E_0 \rangle$ and $\langle w|E_1 \rangle$, but in principle this also depends on the index $w$ if the graph is non-regular. It is worth mentioning that $|\langle w|\tilde{w} \rangle|^2$ represents the success probability $P$, and thus $P$ and $\tau$ are related through $\langle \lambda_1|w \rangle$. 
 
 As the second step of the analysis, we discuss about the optimization of $\gamma$ in Eq. (\ref{Eq:Hamil}) and the selection of the initial state $|\psi(0)\rangle$. The parameter $\gamma$ has to be chosen at an optimal value $\gamma_{opt}$ so that the search will work in the most efficient way. Specifically, $\gamma_{opt}$ is chosen to the value where Eq. (\ref{eq:stateE0}) is approximately true. In our numerical simulation, $\gamma_{opt}$ is determined by finding the point where $|\langle \lambda_1|E_0 \rangle|^2 \approx |\langle \lambda_1|E_1 \rangle|^2 \approx 0.5$ is achieved. This point is shown graphically in Figure \ref{fig:opt}. As well as $\gamma$, the initial state of the time evolution $|\psi(0)\rangle$ has to be chosen properly for the search to work. Clearly the ideal choice is $|\lambda_1\rangle$, since we want the dynamics to stay inside the two-dimensional subspace spanned by $|E_{0,1}\rangle$. On the other hand, the search still works (shows high success probability) using a state that has substantial overlap with $|\lambda_1\rangle$. Here, we utilize the localized property of $|\lambda_1\rangle$ (see the inset of Figure \ref{fig:eigenvector}), and choose an initial state where the quantum walker is fully localized at the single largest degree hub node. We define this state as $|\text{hub}\rangle$ for convenience. On the other hand, we have confirmed that the uniform superposition state over all nodes $\sum_{i=1}^N |i\rangle / \sqrt{N}$, which is conventionally used in spatial search, has small overlap with $|\lambda_1\rangle$. For all of the following results, our simulation is performed with $|\psi(0)\rangle = |\text{hub}\rangle$.

 \begin{figure}[t!]
 \centering
 \subfigure[]{\label{fig:dist25}
  \includegraphics[width=6.2cm]{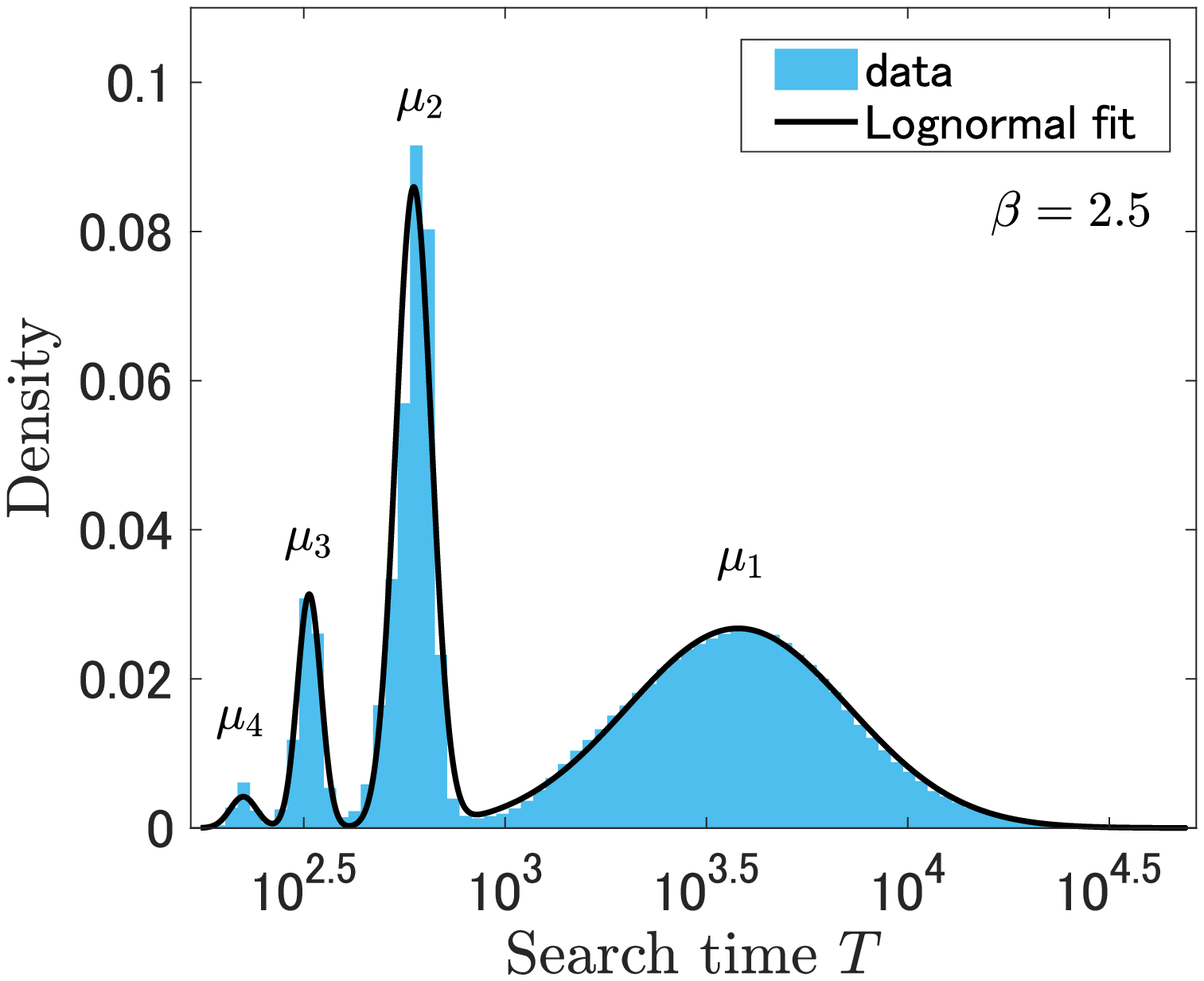}}
 \subfigure[]{\label{fig:dist3}
  \includegraphics[width=6.2cm]{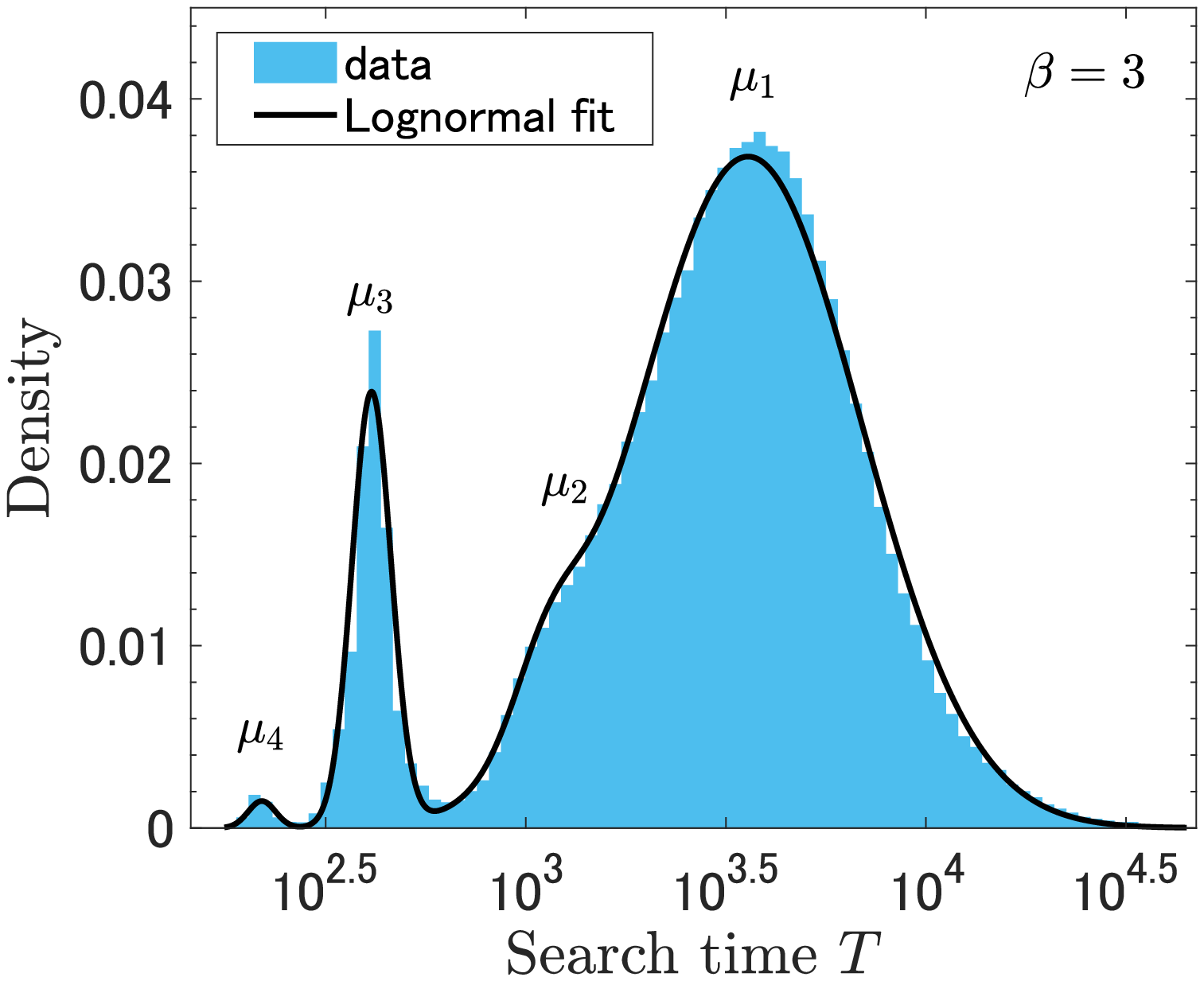}}
 \subfigure[]{\label{fig:dist35}
  \includegraphics[width=6.2cm]{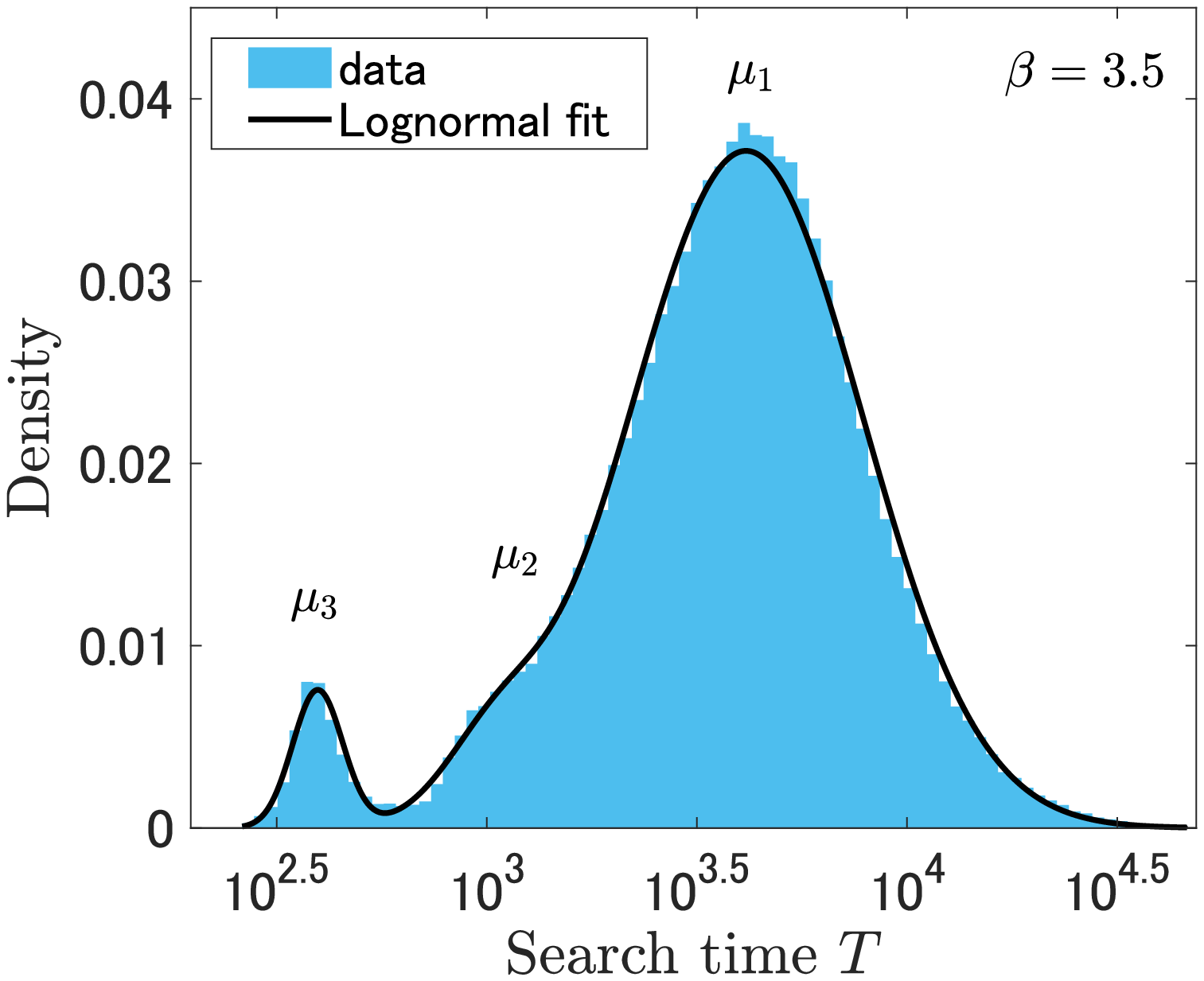}}
  
  \caption{Distributions of the search time $T$, in logarithmic scale, for networks with $N=10000, m=10$ and (a) $\beta=2.5$, (b) $\beta=3$, (c) $\beta=3.5$, respectively. To obtain each distribution we have generated $1000$ network samples and computed $T$ for at least $400$ randomly selected nodes from each network sample. For all distributions we observe multiple peaks, and we find they best fit to a sum of lognormal distributions. The symbols $\mu_1,\cdots,\mu_4$ indicate the individual modes.}
  \label{fig:betadists}
 \end{figure} 
  
 Now let us describe the simulation results, where we have numerically computed the maximum probability to measure the quantum walker after evolving for a optimal time $\tau$, $P(t=\tau) = |\langle w|\exp{({-i} H \tau/\hbar)}|\psi(0) \rangle|^2$. As we know that $\tau$ and $P(\tau)$ depend on the index of the marked node $w$, we will take full account of which node in the network was marked when evaluating the search. To this end, we first show the distribution of the search time $T = \tau/P(\tau)$. We get the distribution by generating multiple samples of the Bollob\'{a}s model with a fixed $\{N,m,\beta\}$, for each network repetitively mark a random node, find $\gamma_{opt}$ and compute $T$, and finally take the histogram of $T$. We have excluded the largest degree hub node when randomly marking a node, since we initialize the quantum walker on that site. Figure \ref{fig:betadists} shows the distribution for three different values of $\beta=2.5,3,3.5$, with fixed $N$ and $m$. Note that the distribution is taken in logarithmic scale. The main feature in this distribution is that they have multiple peaks, meaning that there are classes of nodes that can be searched faster or slower than each other. A good fit to the distributions was a sum of lognormal functions in the form of
 \begin{equation}
 f(T) = \sum_i p_i g(T;\mu_i,\sigma_i),
 \end{equation}
 where
 \begin{equation}
 g(T;\mu,\sigma) = \frac{1}{\sqrt{2\pi} \sigma T} \exp{\left( \frac{(\ln{T} - \mu)^2}{2\sigma^2} \right)}.
 \end{equation}
 Therefore the distribution $f(T)$ is characterized by the mean values $\mu_i$, standard deviations $\sigma_i$ and the mixing parameters $p_i$ (constraint such that $\sum_i p_i = 1$). For the distributions with $\beta=2.5$ and $\beta=3$ we take up to $i=4$, and for $\beta=3.5$ we take up to $i=3$.
 
 We understand that this multi-mode lognormal distribution results from the randomness of the network and the effect of the hub node. When we take a distribution of the search time on the Erd\"{o}s R\'{e}nyi random graph, we see a single-mode lognormal distribution. Likewise, the connections of the nodes in the Bollob\'{a}s model is mostly random (meaning that there are no characteristic structures such as communities or self-similarity) except the overall degree distribution follows a power law. However, this power law degree distribution, or the large degree hub node heavily influences the nodes around it, leading to the multi-mode distribution. In fact, we find that the narrower modes $\mu_3$ and $\mu_4$ (and $\mu_2$ for $\beta=2.5$) corresponds to the nodes that are directly connected to the largest degree hub. We will discuss this further in Section \ref{sec:4}.

  \begin{figure}[t]
 \centering
 \subfigure[]{\label{fig:scaling25}
  \includegraphics[width=6.2cm]{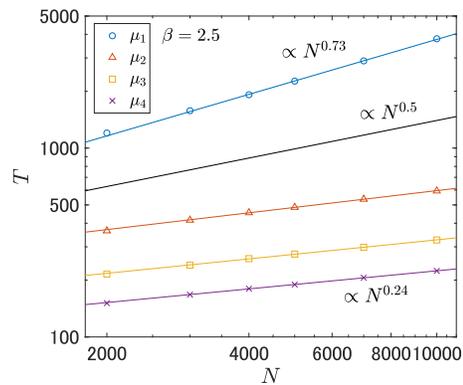}}
 \subfigure[]{\label{fig:scaling3}
  \includegraphics[width=6.2cm]{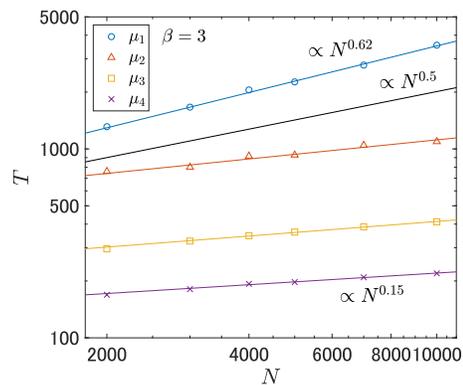}}
 \subfigure[]{\label{fig:scaling35}
  \includegraphics[width=6.2cm]{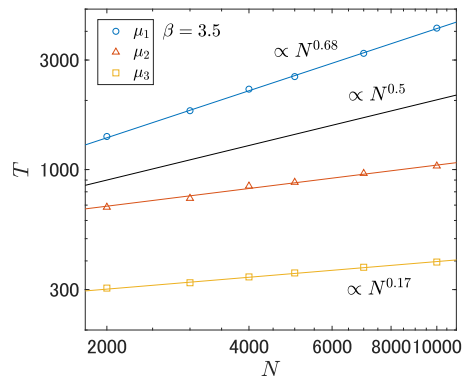}}
 \caption{Dependence on the network size $N$ of the average values of the search time distribution $\mu_1,\cdots,\mu_{4}$ for networks with $m=10$ and (a) $\beta=2.5$, (b) $\beta=3$, (c) $\beta=3.5$, respectively. The black solid line of $N^{0.5}$ is drawn as a reference.} 
 \label{fig:scalingvbeta}
 \end{figure}
 
 \begin{table}[b!]
  \centering
  \renewcommand{\arraystretch}{1.5}
  \setlength{\tabcolsep}{7pt}
   \caption{Exponent $\alpha$ of each parameters of the search time distribution fit to $\propto N^{\alpha}$. Obtained from networks with $N=2000 \sim 10000$, $m=10$ and $\beta=2.5, 3, 3.5$. $const.$ represents that the quantity is independent of $N$.}
   \begin{tabular}{c|c|c|c} \hline
     & $\beta = 2.5$ & $\beta = 3$ & $\beta = 3.5$ \\ \hline
    $\mu_1$ & $0.731 \pm 0.021$ & $0.620 \pm 0.024$ & $0.681 \pm 0.032$ \\
    $\mu_2$ & $0.295 \pm 0.010$ & $0.254 \pm 0.040$ & $0.256 \pm 0.027$ \\
    $\mu_3$ & $0.253 \pm 0.009$ & $0.194 \pm 0.014$ & $0.172 \pm 0.014$ \\
    $\mu_4$ & $0.240 \pm 0.006$ & $0.155 \pm 0.012$ & N/A \\ \hline
    $\sigma_1$ & $0.147 \pm 0.026$ & $0.180 \pm 0.032$ & $0.155 \pm 0.029$ \\
    $\sigma_2$ & $-0.013 \pm 0.005$ & $-0.133 \pm 0.035$ & $const.$ \\
    $\sigma_3$ & $const.$ & $-0.016 \pm 0.004$ & $const.$ \\
    $\sigma_4$ & $const.$ & $const.$ & N/A \\ \hline
    $p_1$ & $0.213 \pm 0.018$ & $0.211 \pm 0.022$ & $0.112 \pm 0.030$ \\
    $p_2$ & $-0.132 \pm 0.034$ & $-1.65 \pm 0.54$ & $-0.942 \pm 0.096$ \\
    $p_3$ & $-0.478 \pm 0.043$ & $-0.423 \pm 0.061$ & $-0.613 \pm 0.057$ \\
    $p_4$ & $-0.847 \pm 0.055$ & $-0.920 \pm 0.086$ & N/A \\ \hline

   \end{tabular}
   \label{tab:scalings}
 \end{table}
   
 Next we evaluate the scaling of the parameters $\mu_i, \sigma_i, p_i$ by examining their dependence on $N$, specifically by fitting to the function $ \propto N^{\alpha}$. The obtained scaling exponents $\alpha$ are shown in Table \ref{tab:scalings}, as well as the plots of $\mu_i$ versus $N$ are shown in Figure \ref{fig:scalingvbeta}. We find two features in our results. First, for all $\beta$, $\mu_1$ has $\alpha > 0.5$ while $\mu_{i \geq 2}$ has $\alpha < 0.5$. As $\alpha=0.5$ is the best known scaling of the spatial search algorithm, the scaling of $\mu_{i \geq 2}$ being $\alpha<0.5$ has to be interpreted carefully, and we are not claiming here that a spatial search faster than $T \propto N^{0.5}$ can be achieved. The search time evaluated here is the case when the measurement of the quantum walker is done at the exact optimal time $\tau$ when the probability $P(\tau)$ maximizes. In order to know the optimal time, one needs to know in prior the properties of the marked node, or at least know that the node is in one the modes of $\mu_{i \geq 2}$ in order to make a reasonable guess of the measurement time. Therefore, our result does not mean a search faster than $N^{0.5}$ can be achieved for some nodes in the network, but rather means the quantum walker can be localized to those nodes quickly. Additionally, by identifying the number of nodes $\tilde{N}$ that is involved in the modes $\mu_{i \geq 2}$, and by fitting to $\tilde{N}^\alpha$, the scaling reduces to $\alpha \approx 0.5$.
 
 The second feature in our result is the agreement between the scaling of $p_i$ and the property of the network. From Table \ref{tab:scalings}, we see that $p_{i \geq 2}$ decays as $N$ grows. This corresponds to the decay of the fraction of nodes those are neighbours the largest hub node, $N^{1/(\beta-1)}/N$. The result suggests that the modes $\mu_{i \geq 2}$ corresponds to the nodes those are neighbouring to the hub, or the nodes heavily influenced by the hub. This argument is also supported by the change of the distributions depending on $\beta$ (see Figure \ref{fig:betadists}). As $\beta$ increases, edges will be less concentrated on the large degree nodes, letting the network to become closer to a random graph. This effect is observed as the shrinking of the $\mu_{i \geq 2}$ modes when $\beta$ increases. We note that these scaling obtained from numerical simulations are only guaranteed for $N=2000 \sim 10000$, the region where we executed the simulations.
 
 As a conclusion of this section, the distribution of the search time $T$ obtained by marking different nodes in the network strictly reflects the structure of the network; the randomness and scale-free property (i.e. existence of the hub) leads to a multi-mode lognormal distribution of $T$. The existence of the hub allows the quantum walker to localize on nodes that are neighbours of the hub especially fast.

\section{\label{sec:4}Characterization of the search through network centrality measures}
 In this section, we interpret the search time $T$ and the dynamics of the spatial search by investigating some centrality measures of the network. This will bridge the knowledge in complex network science and quantum dynamics. We investigate the correlation between the search time and six different centrality measures: degree centrality, eigenvector centrality \cite{eigencent}, closeness centrality \cite{closecent}, betweenness centrality \cite{betcent}, random walk closeness centrality \cite{rwclose} and random walk betweenness centrality \cite{rwbet}. The essential result we show here is that the search time is dependent on how close the marked node is to all other node in the network, in terms of shortest path distances.
 
 \begin{figure*}[htp]
 \centering
 \subfigure[]{\label{fig:degree}
  \includegraphics[width=5.8cm]{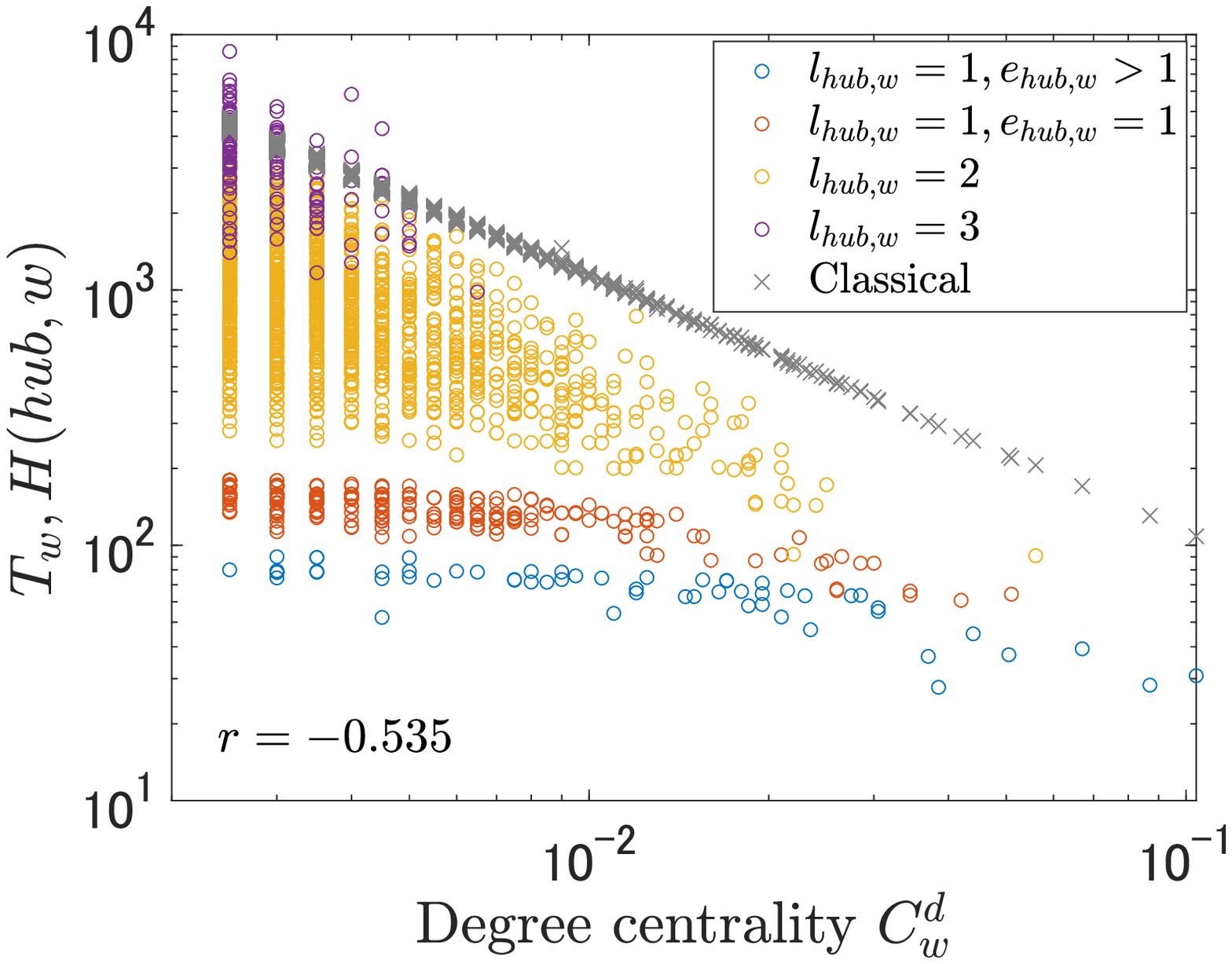}}
 \subfigure[]{\label{fig:eigen}
  \includegraphics[width=5.8cm]{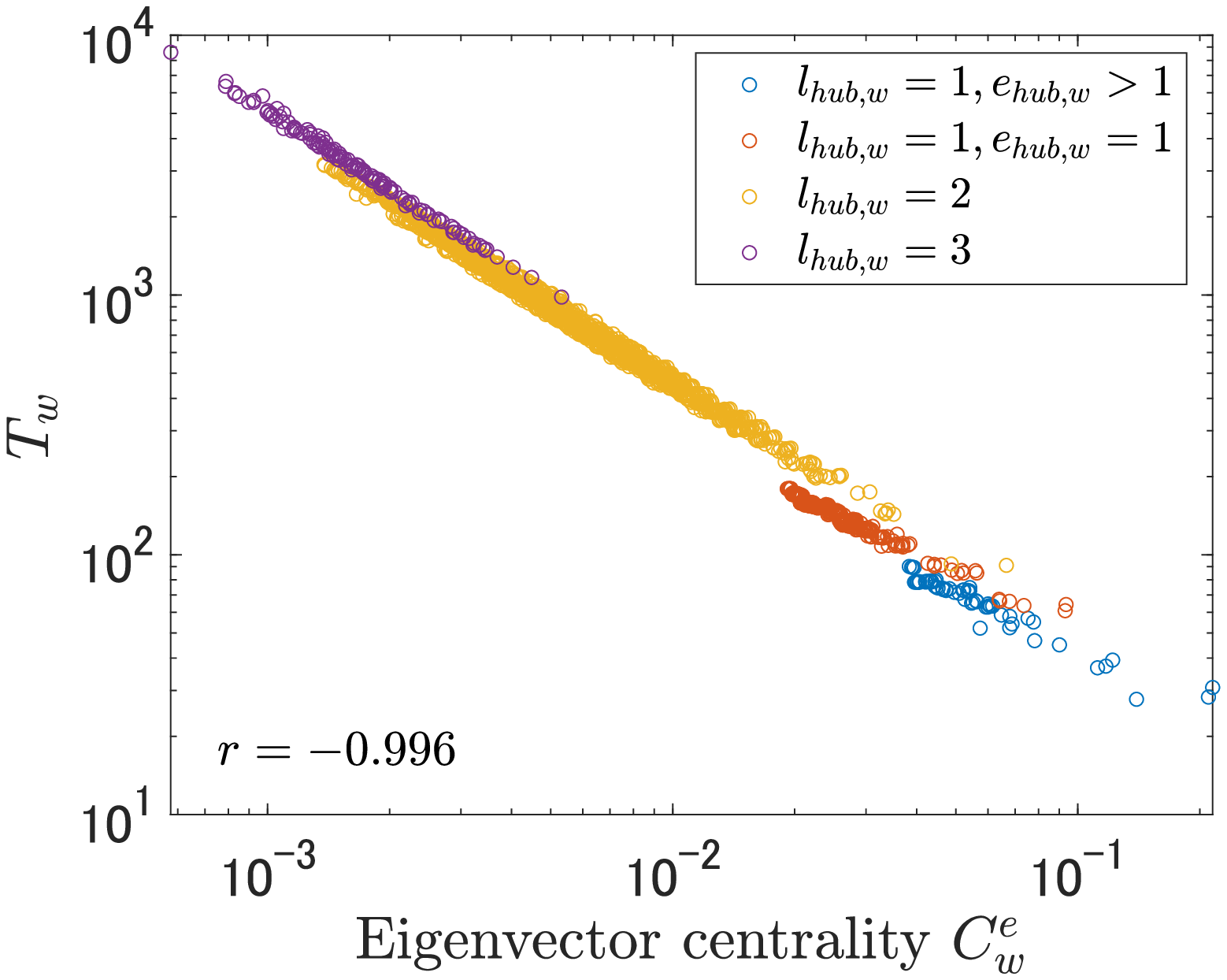}}
 \subfigure[]{\label{fig:closeness}
  \includegraphics[width=5.8cm]{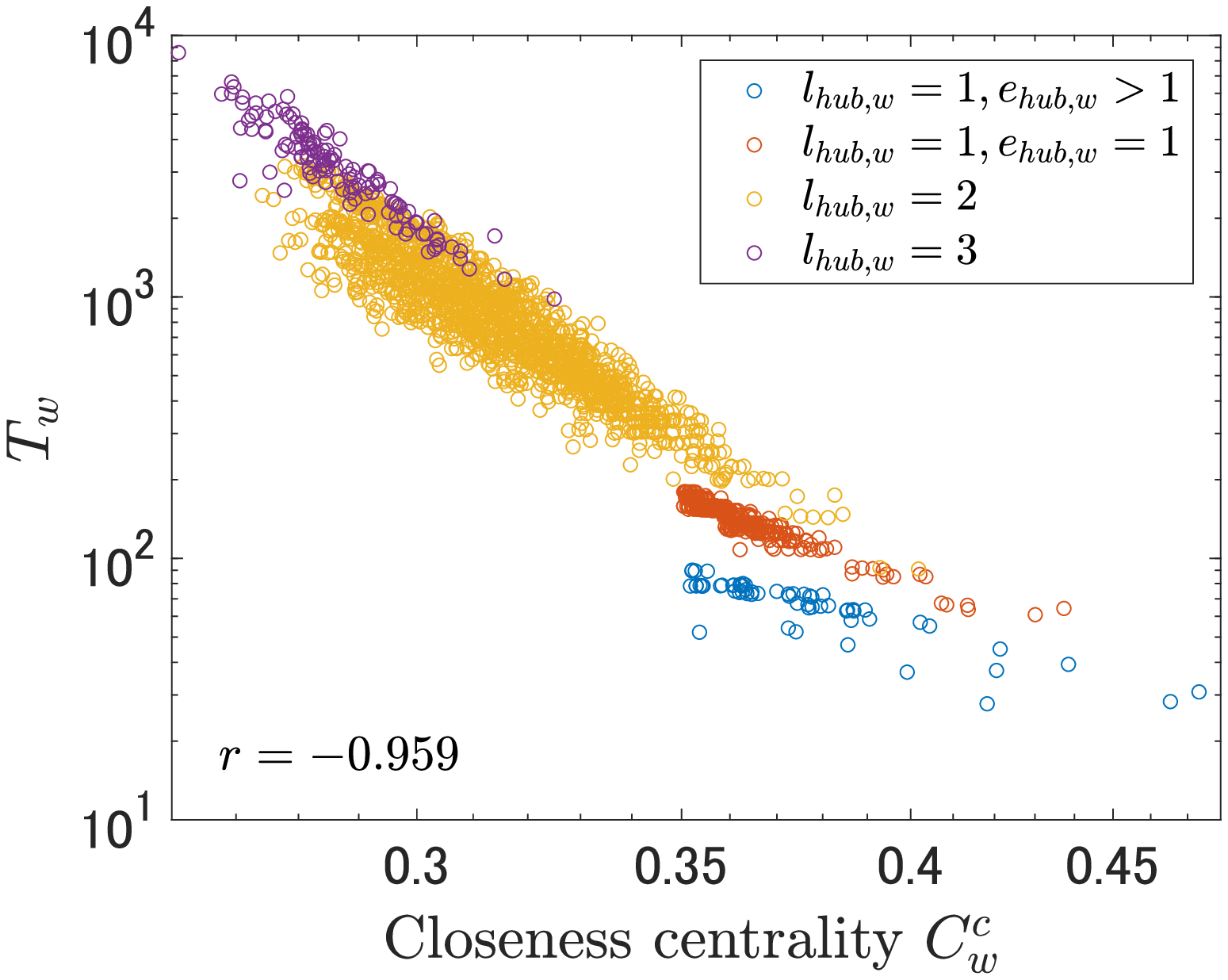}}
 \subfigure[]{\label{fig:betweenness}
  \includegraphics[width=5.8cm]{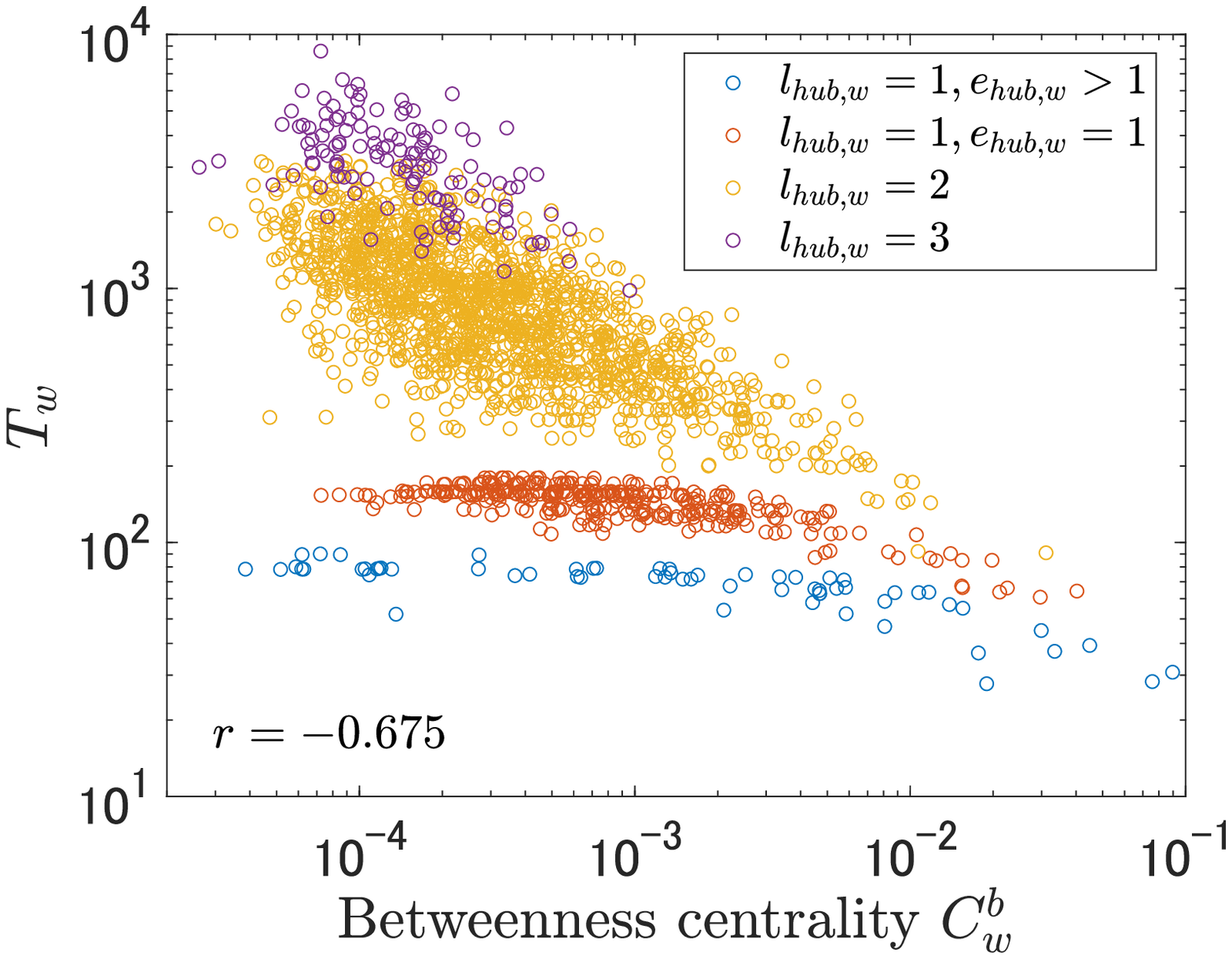}}
 \subfigure[]{\label{fig:rwcloseness}
  \includegraphics[width=5.8cm]{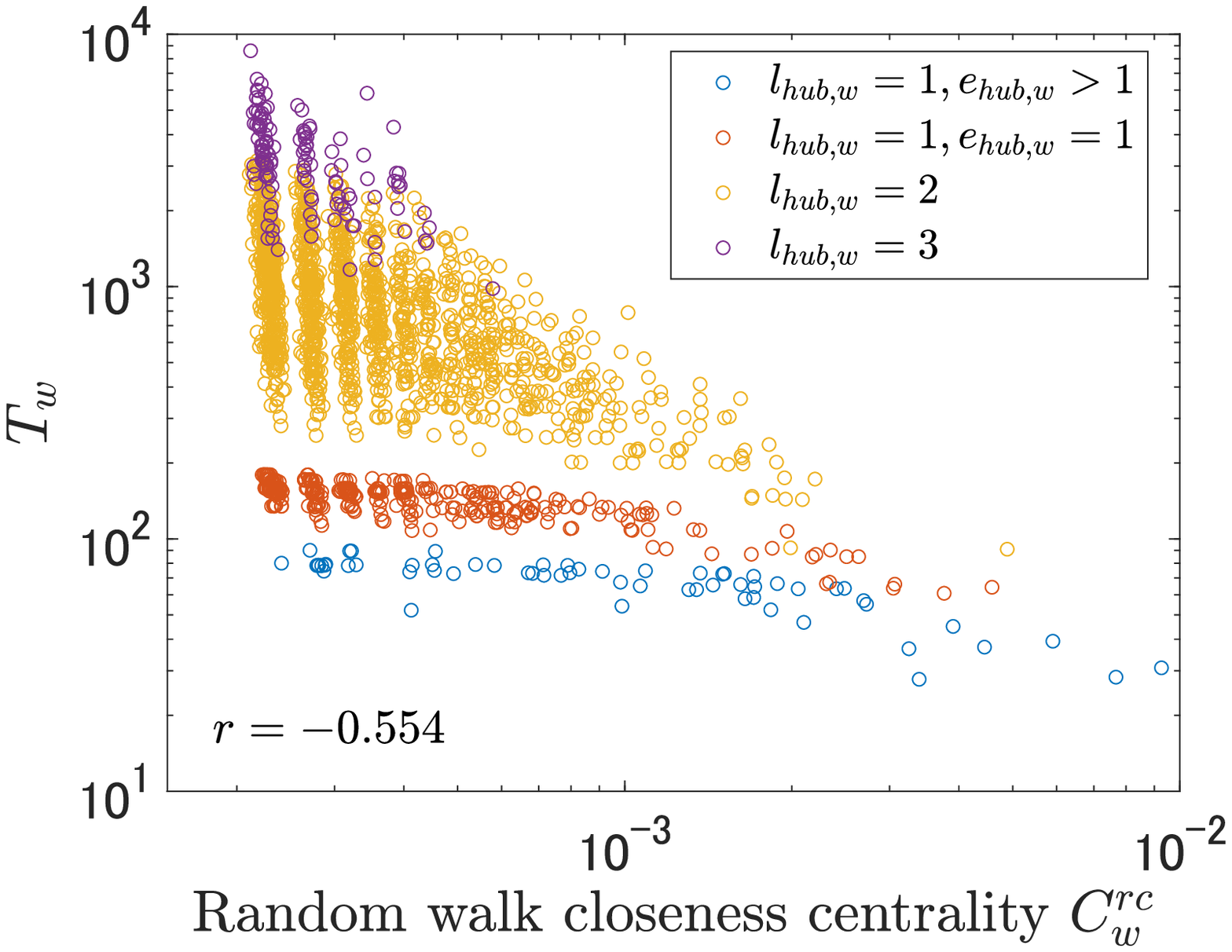}}
 \subfigure[]{\label{fig:rwbetweenness}
  \includegraphics[width=5.8cm]{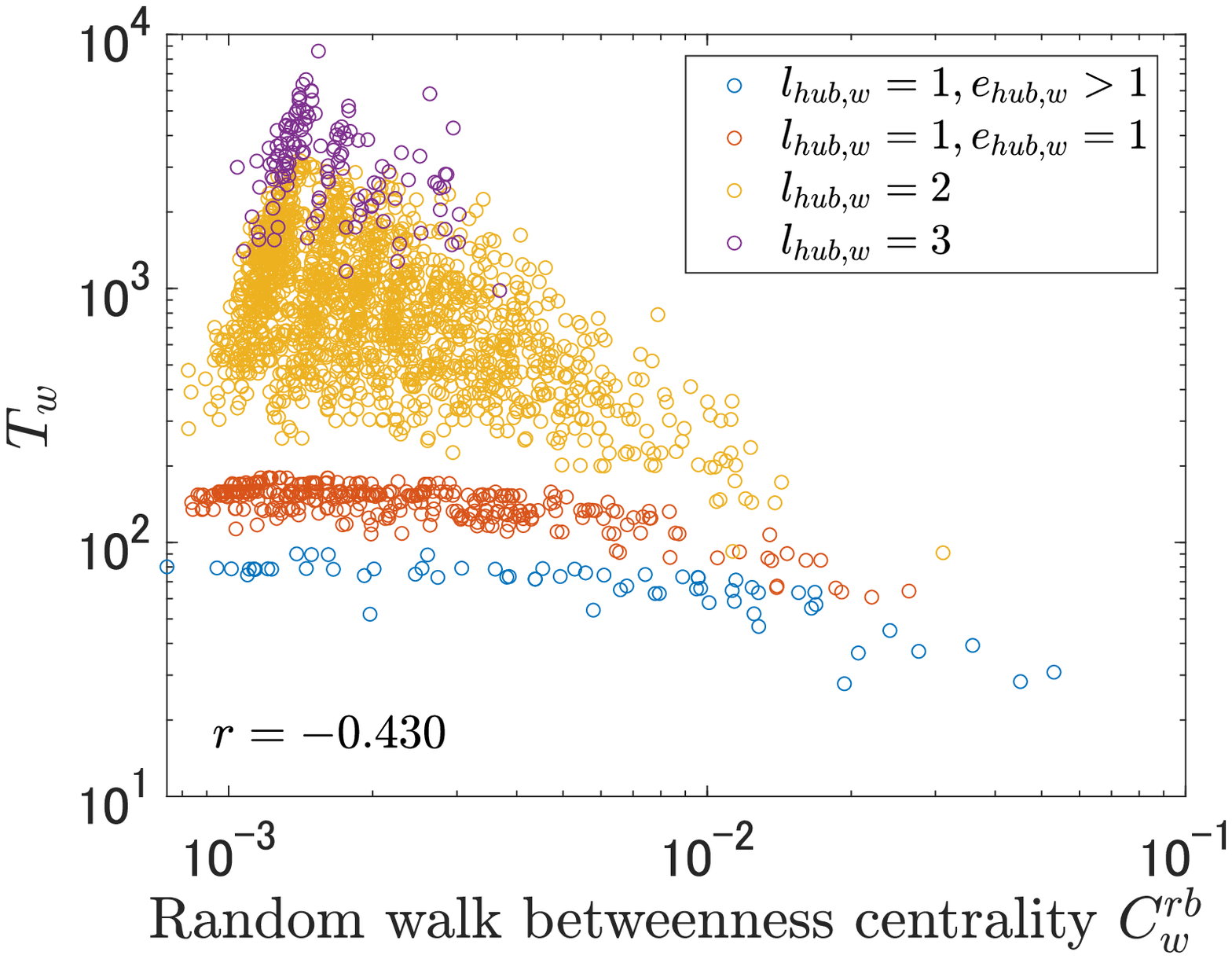}}
  
  \caption{The correlation between the search time $T$ and six different network centrality measures, which are (a) degree centrality, (b) eigenvector centrality, (c) closeness centrality, (d) betweenness centrality, (e) random walk closeness centrality and (f) random walk betweenness centrality. All of the scatter plots are in log-log scale, and each data points corresponds to each node of the network with parameters $N=2000, m=5, \beta=3$. Data points are coloured depending on the shortest path distance between the largest degree hub node and the marked node $l_{hub,w}$, as well as depending on whether the marked node has parallel edges to the hub or not for $l_{hub,w}=1$. $e_{hub,w}$ represents the number of parallel edges between the hub and the marked node (or equivalently the weight $A_{hub,w}$). In (a), the mean first passage time of the classical random walk $H(hub,w)$ (a walk from the hub to the marked node $w$) is also plotted. The best correlation between the centrality measures and the quantum search time is seen from (b) eigenvector centrality, followed by (c) closeness centrality. The Pearson correlation coefficient $r$ of the centrality measures and the search time (both in logarithmic scale) is shown inside the figures.}
  \label{fig:centrality}
 \end{figure*}
 
 The scatter plots where the search time $T$ is plotted against different centrality measures are shown in Figure \ref{fig:centrality}. Figure \ref{fig:degree} shows the case of degree centrality $C^d_w = \sum_{j}^N A_{wj} / (N-1)$, which measures the fraction of nodes that are connected to the node $w$. This plot tells that when the marked node has large degree, the quantum walker will likely to be localized on that node quickly, but if the node has low degree, the search time is almost independent of the degree. This feature is quite different from the case of classical random walk, where the difference can be seen by comparing with the mean first passage time $H(hub,w)$ computed numerically and plotted in the same figure. The mean first passage time $H(i,j)$ is defined as the average time for the classical random walker to visit node $j$ for the first time, starting the walk from node $i$. We can interpret $H(hub,w)$ as the average time to search the marked node by starting the classical random walk from the hub node. From Figure \ref{fig:degree}, we can confirm that the time it takes to search a node using random walk is proportional to its degree \cite{Noh2004}, as well as revealing that the search using quantum walk clearly shows a different feature.
 
 Figure \ref{fig:eigen} shows the case of eigenvector centrality $C^e_w = |\langle \lambda_1|w \rangle|$, which is a centrality measure based on the leading eigenvector of the adjacency matrix. The plot shows a high correlation as expected from Eq. (\ref{eq:energygap}). We also see a good correlation in Figure \ref{fig:closeness} which shows the case of closeness centrality $C^c_w = (N-1) / \sum_{j\neq w}^{N-1} l_{wj}$ where $l_{wj}$ is the shortest path distance between nodes $w$ and $j$. This measure represents how fast one can move from the node $w$ to all other nodes using the shortest paths. 
 
 Figure \ref{fig:betweenness} shows the betweenness centrality $C^b_w = \sum_{i\neq w\neq j} \sigma_{ij}(w) / \sigma_{ij}$, where $\sigma_{ij}$ is the number of shortest paths from node $i$ to $j$, and $\sigma_{ij}(w)$ is the number of shortest paths that goes through node $w$ among them. The random walk closeness centrality $C^{rc}_w = N/\sum_{j} H(j,w)$ in Figure \ref{fig:rwcloseness} is an alternative measure of the closeness centrality, where path lengths between nodes are measured based on the random walk process. The random walk betweenness centrality $C^{rb}_w$ in Figure \ref{fig:rwbetweenness} is an alternative measure of betweenness centrality, where instead of counting only shortest paths, all paths contribute to the measure with a certain weight. All three of these measures are correlated with the search time in a similar way as the degree centrality.
 
The results presented in Figure \ref{fig:centrality} tells us that the quantum walk or the spatial search is a dynamics relying on the shortest paths of the network, unlike classical random walk. In the case of classical random walk, the walker chooses one neighbour randomly at each time step, and thus it is natural to understand that a node having larger degree will have higher probability to receive the walker, leading to shorter time of the search. In contrast, since the quantum walker spreads to all of the neighbours as a superposition state, the length of shortest paths between the nodes determines the time for the complex amplitudes to reach from a node to another, rather than the degrees. As indicated by the high correlation to the closeness centrality $C^c_w$, if the marked node $w$ is averagely close to all other nodes (i.e. has high $C^c_w$), the quantum walker can localize on that node faster since the complex amplitudes of the quantum walker can be collected from the whole network with a shorter time.
 
 The importance of the distances is emphasized by distinguishing the data points in Figure \ref{fig:centrality} based on the shortest path distance between the hub node and the marked node (see the legend of the figure). The data is well clustered depending on $l_{hub,w}$. Especially when the marked node is adjacent to the hub ($l_{hub,w}=1$), these nodes have small shortest path distances with the other nodes by going through the hub, leading to the shortness of $T$. 
 
We also examined the how the scaling of the search time $T \propto N^{\alpha}$ depends on the distance between the hub and the marked node $l_{hub,w}$. We computed multiple samples of $T_w$ from network with parameters $N=2000 \sim 10000, m=5, \beta=3$ and took the averaged of $T_w$ for each $l_{hub,w}$. The obtained scaling $\alpha$ is shown in Table \ref{tab:scaling_by_distance}. Although we get large standard deviations of $T_w$ since the factor determining the search time is not only $l_{hub,w}$, the scaling $\alpha$ roughly increases linearly as $l_{hub,w}$ grows.
 
  \begin{table}[b!]
    
  \centering
  \renewcommand{\arraystretch}{1.5}
  \setlength{\tabcolsep}{7pt}
   \caption{Exponent $\alpha$ of the average search time $T \propto N^{\alpha}$ for nodes with different distances from the hub $l_{hub,w}$. Networks with parameters $N=2000 \sim 10000$, $m=5$, $\beta=3$ are used to obtain $\alpha$. }   
   \begin{tabular}{c|c|c|c} \hline
    $l_{hub,w}$ & $1$ & $2$ & $3$ \\ \hline
    $\alpha$ & $0.120 \pm 0.019$ & $0.638 \pm 0.122$ & $1.127 \pm 0.205$ \\ \hline
    
   \end{tabular}
   
   \label{tab:scaling_by_distance}
 \end{table}
 
 Note that the especially short $T$ when the marked node is adjacent to the hub is not due to the localized initial state of the quantum walker. The quantum walker does not instantaneously hop from the hub to the marked node, but instead has to traverse the entire network and acquire some phase to localize on the marked node. In fact, from Eq. (\ref{eq:energygap}) we can see that the optimal evolution time $\tau = \pi / \Delta E$ is independent of the initial state. The initial state determines the fraction of the complex amplitude that stays in the two-dimensional subspace spanned by $|E_0\rangle$ and $|E_1\rangle$, and thus only affects the maximum success probability $P(\tau)$. 
 
 Although the high correlation between the search time and the eigenvector centrality is expected from Eq. (\ref{eq:energygap}), there are small corrections from the factor $|\langle w|\tilde{w} \rangle|$, which is essentially the success probability $P$. In our results, we did not see a particularly high correlation between the centrality measures and $P$. The best correlation we could observe was with the eigenvector centrality, with correlation coefficient $r = 0.363$.

\section{\label{sec:5}Discussions}
In this paper, we have analyzed the performance of the continuous-time spatial search algorithm on the Bollob\'{a}s model, which is a scale-free network. We found that the search time is faster as the marked node is more central in the network, where this is measured by the closeness centrality of the node. Such feature results from the power law degree distribution and the existence of the large degree hub node of the scale-free network. Interestingly, the degree of the marked node does not crucially matter for the search time, but the shortest path distances between the marked node and the rest of the nodes rather determines the search time. We can interpret that the search time is dependent on how fast the marked node can collect the complex amplitudes globally from the network (and thus the global structure matters). This is in contrast to searching by classical random walk which highly depends on how many edges are locally connected to the marked node. We also observed that the distribution of the search time in a network follows a multi-mode lognormal distribution, which well reflects the structure of the scale-free network. We achieved to characterize the interesting relationship between the network structure and the performance of the spatial search algorithm, which could not have been discovered using regular or homogeneous graphs.

The localized property of the leading eigenvector of the adjacency matrix was advantageous in a way that we could select a initial state fully localized on a single node, instead of a superposition state. We can generally say that if the search Hamiltonian Eq. (\ref{Eq:Hamil}) without the $\epsilon_w|w\rangle\langle w|$ term has a localized ground state, one can choose a localized initial state. This may be advantages in experimental implementations, since preparing a superposition state with arbitrary amplitudes and relative phases can be difficult \cite{singlemagnet}. However, such localized leading eigenvector also creates differences in the optimal measurement time $\tau$ depending on the marked node. This fundamentally limits the ability to perform the spatial search algorithm, since we need some amount of information about the node that is searched for in order to estimate the measurement time. However, as the distribution of the search time is well separated into classes depending on the node being adjacent to the hub or not, on can make a reasonable guess of $\tau$ by limiting the nodes to mark within one class. In addition, we can naturally translate the dynamics of spatial search algorithm into a efficient state transfer protocol between the hub node and a single marked node \cite{Chak1,transfer1}, which is a simple and useful application.

\begin{acknowledgements}
We thank Benjamin Renoust for helpful discussions. This project was made possible through the support of a grant from the John Templeton Foundation, the grant ID 60478. The opinions expressed in this publication are those of the authors and do not necessarily reflect the views of the John Templeton Foundation. This work was also supported in part by from the Japanese program Q-LEAP. BC and YO thank the support from Funda\c{c}\~{a}o para a Ci\^{e}ncia e a Tecnologia (Portugal), namely through programme POCH and projects UID/EEA/50008/2013 and UID/EEA/50008/2019, as well as from the project TheBlinQC supported by the EU H2020 QuantERA ERA-NET Cofund in Quantum Technologies and by FCT (QuantERA/0001/2017), and from the EU H2020 Quantum Flagship projects QIA (820445) and QMiCS (820505). 
\end{acknowledgements}

\end{document}